\documentclass[aps,pra,twocolumn,eqsecnum,superscriptaddress,amsmath,amssymb,showpacs]{revtex4}
\usepackage{bm}

\begin{document}
% --------------------  TITLE  --------------------
\title{Solution of the Lindblad Equation in the Kraus Representation}

% ------------  AUTHORS AND AFFILIATIONS ----------
\author{H. Nakazato}\email{hiromici@waseda.jp}
\author{Y. Hida}
\affiliation{Department of Physics, Waseda University, Tokyo 169-8555, Japan}

\author{K. Yuasa}\email{kazuya.yuasa@aist.go.jp}
\affiliation{Research Center for Information Security, National Institute of Advanced Industrial Science and Technology (AIST), 1-18-13 Sotokanda, Chiyoda-ku, Tokyo 101-0021, Japan}

\author{B. Militello}
\author{A. Napoli}
\author{A. Messina}
\affiliation{MIUR and Dipartimento di Scienze Fisiche ed
Astronomiche dell'Universit\`{a} di Palermo, Via Archirafi 36,
I-90123 Palermo, Italy}

\date[]{November 8, 2006}

% --------------------  ABSTRACT  --------------------
\begin{abstract}
The so-called Lindblad equation, a typical master equation describing the
dissipative quantum dynamics, is shown to be solvable for
finite-level systems in a compact form without resort to writing
it down as a set of equations among matrix elements. The solution
is then naturally given in an operator form, known as the Kraus
representation. Following a few simple examples, the general
applicability of the method is clarified.
\end{abstract}
\maketitle

% --------------------  INTRODUCTION  ------------------
\section{Introduction}
\label{sec:intro}
A quantum system in interaction with an environment, that is a larger system with a continuous number of degrees of freedom, displays an irreversible, dissipative, and decohering dynamics, which is described by a master equation \cite{Weiss,ref:MilburnTextbook,ref:KuboTextbook,ref:ScullyTextbook,ref:CarmichaelTextbook,Alicki,Petruccione,Gardiner}.
The ability of solving the master equations plays a considerable role in different physical contexts both from practical and fundamental points of view.
It enables us to discuss the thermalization processes and to highlight the problem of individualizing the quantum--classical boundary \cite{Zurek}.
Master equations are also necessary for the realistic description of many quantum systems exploited in quantum information processing \cite{qiproc,ref:NielsenChuang,ref:BouwmeesterZeilingerEkert,ref:BennettDiVincenzoReview,ref:Haroche,ref:JJqubitRMP,ref:MartinDelgadoReview2002,ref:VandersypenRMP}.

The so-called Lindblad equation \cite{Lindblad} is the most general form of linear and Markovian master equation
preserving the probability and the positivity of the density
operator.
Although the Lindblad equation can be derived, under appropriate approximations or in some limits, from a microscopic model consisting of the quantum system and the environment after tracing out the latter degrees of freedom \cite{Weiss,ref:MilburnTextbook,ref:KuboTextbook,ref:ScullyTextbook,ref:CarmichaelTextbook,Alicki,Petruccione,Gardiner}, one can even start with it as a phenomenological equation \cite{pheno}.

Generally speaking, solving master equations requires the resolution of a set of differential equations among the matrix elements of the density operator with respect to a specific basis, usually the eigenstates of the system Hamiltonian.
Nevertheless, it is generally a cumbersome task to list up all of the $N^2$ equations, $N$ being the dimension of the density matrix (the number of states), and if one selects the basis carelessly, many of the equations look coupled to each other, even though they are not essentially so on the appropriate basis.

Moreover, even in those lucky cases wherein one can find the solution to the set of the differential equations, the density operator of the system is given as a collection of the matrix elements with respect to the specific basis, instead of being expressed in an operator form.
It is known that the evolution of the density operator is given in the Kraus representation \cite{Kraus}, but it is not easy to construct it from the set of the matrix
elements.
Over the last decade, several efforts have been done in this direction and several procedures for constructing operator solutions to master equations under appropriate conditions have been proposed \cite{LDBDeq_sol,LDBDeq_sol6}.

In this article, we present a procedure for solving Markovian master equations directly in an operator form in the Kraus representation, instead of giving it by specifying its matrix elements.
If one works with the matrix elements, one needs to write down an $N^2\times N^2$ matrix which couples the $N^2$ equations for the matrix elements, and then, realizes that many of them are not really coupled to each other (for relatively simple systems).
In the approach discussed here, on the other hand, it is not required to write such a sparse matrix, where most of its elements are zero, but one can realize the minimal set of the necessary equations by looking at the relevant part of the dissipator of the master equation, associated with the ``quantum jumps'' \cite{QuantumJump,QuantumJump2}.

The approach is initially sketched in Sec.~\ref{sec:example} with two simple examples, few-level systems at zero and finite temperatures, showing the idea and its powerfulness.
Then, we discuss its general applicability in Sec.~\ref{sec:general} and conclude the article in Sec.~\ref{sec:conclusion}\@.
In Appendices~\ref{app:DerivationMasterEq} and \ref{app:SolModel2_HN}, we summarize a derivation of master equations, giving possible connections between the master equations discussed in the main text and microscopic Hamiltonians, and present some details about the solution to the second example considered in Sec.~\ref{sec:Model2}\@.

\section{Simple Examples}
\label{sec:example}
In this section, the resolution of two master equations describing
the dynamics of two physical systems is presented. The first
example we discuss refers to a thermal reservoir at zero temperature
whereas in the second one a finite-temperature reservoir ($T\not=0$) is
considered. We successfully solve the respective master equations
building up in both cases solutions in the operator form known as
the Kraus representation instead of expressing them in terms of a
collection of matrix elements. These two examples aim at
illustrating the ``key ingredients'' and the basic steps of
the general procedure we are going to present in Sec.~\ref{sec:general}\@.

\subsection{A Three-Level System at the Zero Temperature}
\label{sec:Model1}
Consider a three-level system at the zero temperature and suppose its dynamics is described by the following master equation
\begin{align}
\dot{\rho}(t)
={}&{-i}[H,\rho(t)]\nonumber\\
&{}-\gamma\left(
\frac{1}{2}\{|1\rangle\langle1|,\rho(t)\}
-|0\rangle\langle1|\rho(t)|1\rangle\langle0|
\right)\nonumber\\
&{}-\gamma\left(
\frac{1}{2}\{|2\rangle\langle2|,\rho(t)\}
-|1\rangle\langle2|\rho(t)|2\rangle\langle1|
\right),\label{eq:ldbdeq3}
\end{align}
where the states $|i\rangle$ ($i=0,1,2$) diagonalize the Hamiltonian $H$,
\begin{equation}
H=\sum_{i=0,1,2}E_i|i\rangle\langle i|.
\label{eqn:SpectrumModel1}
\end{equation}
This master equation can describe the dissipative dynamics in the
triplet sector of two qubits at the zero temperature, as shown in
Appendix~\ref{app:Model1}\@. In order to solve the master equation
(\ref{eq:ldbdeq3}), observe that it can be rewritten in the following form:
\begin{equation}
\dot{\rho}(t)
=A\rho(t)+\rho(t)A^\dag+\gamma\sum_{i=0,1}B_i\rho(t)B_i^\dag
\label{eq:L1}
\end{equation}
with
\begin{subequations}
\begin{gather}
A=-iH-\frac{\gamma}{2}\,\Bigl(
|1\rangle\langle1|+|2\rangle\langle2|
\Bigr),\\
B_0=|0\rangle\langle1|,\quad
B_1=|1\rangle\langle2|.
\end{gather}
\end{subequations}
This rewriting of the master equation naturally suggests one to
introduce the transformation
\begin{equation}
\rho(t)=e^{At}\rho_\text{I}(t)e^{A^\dag t}.
\label{eqn:IntPic}
\end{equation}
It is indeed easy to convince oneself that $\rho_\text{I}(t)$
evolves according to the equation
\begin{equation}
\dot{\rho}_\text{I}(t)
=\gamma\sum_{i=0,1}B_i(t)\rho_\text{I}(t)B_i^\dag(t),\quad
B_i(t)=e^{-At}B_ie^{At}.
\label{eqn:LindbladEqInt}
\end{equation}
On the other hand, explicitly evaluating the operators $B_i(t)$,
we obtain
\begin{equation}
B_0(t)=B_0e^{-i(E_1-E_0)t}e^{-\gamma t/2},\quad
B_1(t)=B_1e^{-i(E_2-E_1)t},
\end{equation}
and then (\ref{eqn:LindbladEqInt}) is expressed as
\begin{equation}
\dot{\rho}_\text{I}(t)
=\gamma B_0\rho_\text{I}(t)B_0^\dag e^{-\gamma t}
+\gamma B_1\rho_\text{I}(t)B_1^\dag.
\label{eqn:RhoIEq}
\end{equation}

It is worth noting that $B_i$'s are essentially ladder operators
satisfying the following relations,
\begin{equation}
B_0^2=B_1^2=B_1B_0=0,\quad B_0B_1=|0\rangle\langle2|\neq0,
\end{equation}
so that we get
\begin{equation}
B_0\dot{\rho}_\text{I}(t)B_0^\dag=\gamma
B_0B_1\rho_\text{I}(t)B_1^\dag B_0^\dag, \quad
B_1\dot{\rho}_\text{I}(t)B_1^\dag=0.
\label{eqn:ElementRhoIEq}
\end{equation}
These equations are easily integrated to obtain
\begin{subequations}
\label{eqn:ElementRhoI}
\begin{align}
B_1\rho_\text{I}(t)B_1^\dag
&=B_1\rho_\text{I}(0)B_1^\dag
=B_1\rho(0)B_1^\dag,\\
B_0\rho_\text{I}(t)B_0^\dag
&=B_0\rho(0)B_0^\dag
+\gamma tB_0B_1\rho(0)B_1^\dag B_0^\dag.
\end{align}
\end{subequations}
It is now straightforward to deduce the expression of
$\rho_\text{I}(t)$, by simply integrating the right-hand side of
(\ref{eqn:RhoIEq}) with (\ref{eqn:ElementRhoI}) taken into
account,
\begin{multline}
\rho_\text{I}(t)
=\rho(0)
+(1-e^{-\gamma t})B_0\rho(0)B_0^\dag
+\gamma tB_1\rho(0)B_1^\dag\\
+(1-e^{-\gamma t}-\gamma te^{-\gamma t})
B_0B_1\rho(0)B_1^\dag B_0^\dag.
\label{eqn:RhoIModel1}
\end{multline}
Finally, coming back to the original picture through the operator
\begin{multline}
e^{At}=e^{-iE_2t}e^{-\gamma t/2}|2\rangle\langle2|
+e^{-iE_1t}e^{-\gamma t/2}|1\rangle\langle1|\\
+e^{-iE_0t}|0\rangle\langle0|,
\end{multline}
we get the explicit expression of the density operator at an arbitrary time $t>0$,
\begin{align}
\rho(t)
={}&e^{At}\rho(0)e^{A^\dag t}\nonumber\\
&+(1-e^{-\gamma t})B_0\rho(0)B_0^\dag
+\gamma te^{-\gamma t}B_1\rho(0)B_1^\dag\nonumber\\
&+(1-e^{-\gamma t}-\gamma te^{-\gamma t})
B_0B_1\rho(0)B_1^\dag B_0^\dag.
\label{eq:sol1}
\end{align}

We wish to stress that the master equation has been solved in the
operator form and its solution naturally falls into the Kraus
representation. As expected, the procedure presented in this
section does not take into account at all the set of the equations
among the matrix elements, which, in this case, involves a
$9\times9$ matrix (or at least, two equations for the diagonal
elements and three equations for the off-diagonal elements). We
only need two coupled equations (\ref{eqn:ElementRhoIEq}).
Therefore, a great simplification has been achieved through the
present approach. A key element is to introduce the transformation
(\ref{eqn:IntPic}) and rewrite the master equation into the form
(\ref{eqn:LindbladEqInt}). The number of the necessary equations
(two in this case) relies on the structure of the right-hand side
of (\ref{eqn:LindbladEqInt}), related to the so-called ``quantum
jumps'' \cite{QuantumJump,QuantumJump2}.

In the case under scrutiny, the thermal reservoir is at the zero temperature, which results in the vanishing derivative in the second
equation in (\ref{eqn:ElementRhoIEq}). This feature makes easier the resolution of the set of the equations in (\ref{eqn:ElementRhoIEq}):
in fact, one can proceed one by one, solve the second equation
first and then the first one after plugging the solution of the
second into the first. We stress that this feature is special for
zero-temperature cases and reflects the fact that no transitions
into the higher energy levels from the lower energy ones occur at the zero
temperature.
Similar structures are found in Refs.~\cite{LDBDeq_sol6,QuantumJump2}.

\subsection{A Two-Level System at a Finite Temperature}
\label{sec:Model2}
The second example is a single two-level system at a finite temperature, described by the master equation
\begin{align}
\dot{\rho}(t)
={}&{-\frac{i}{2}}\Omega[\sigma_z,\rho(t)]\nonumber\\
&{}-\gamma_+\left( \frac{1}{2}\{\sigma_+\sigma_-,\rho(t)\}
-\sigma_-\rho(t)\sigma_+ \right)\nonumber\\
&{}-\gamma_-\left(
\frac{1}{2}\{\sigma_-\sigma_+,\rho(t)\} -\sigma_+\rho(t)\sigma_-
\right). \label{eqn:Model2-0}
\end{align}
$\sigma_z$ is a spin operator and $\sigma_\pm$ are ladder operators.
The first term on the right-hand side describes the unitary evolution, the second the spontaneous and the induced decays from the upper to the lower level, and the third the induced transition from the lower to the upper.
This master equation is derived from a spin-boson model: see Appendix~\ref{app:Model2}\@.

As in the previous example, we arrange the master equation into
\begin{equation}
\dot{\rho}_\text{I}(t)
=\gamma_+\sigma_-\rho_\text{I}(t)\sigma_+e^{-\gamma t}
+\gamma_-\sigma_+\rho_\text{I}(t)\sigma_-e^{\gamma t},
\label{eqn:MasterEqB-0}
\end{equation}
where $\rho_\text{I}(t)$ is defined by (\ref{eqn:IntPic}) with
\begin{subequations}
\begin{gather}
A=-\frac{1}{4}\gamma^\beta
-\frac{1}{4}(\gamma+2i\Omega)\sigma_z,\\
\gamma^\beta=\gamma_++\gamma_-,\quad
\gamma=\gamma_+-\gamma_-.
\end{gather}
\end{subequations}
Starting from (\ref{eqn:MasterEqB-0}) and taking into account
that $\hat{\sigma}_{\pm}^{2}=0$, it is immediate to reach the
following equations:
\begin{equation}
\begin{cases}
\medskip
\displaystyle
\sigma_+\dot{\rho}_\text{I}(t)\sigma_-
=\gamma_+\sigma_+\sigma_-\rho_\text{I}(t)\sigma_+\sigma_-e^{-\gamma t},\\
\displaystyle
\sigma_-\dot{\rho}_\text{I}(t)\sigma_+
=\gamma_-\sigma_-\sigma_+\rho_\text{I}(t)\sigma_-\sigma_+e^{\gamma t}.
\end{cases}
\label{eqn:Projected-0}
\end{equation}
In contrast to the previous example,
i.e.~(\ref{eqn:ElementRhoIEq}) for the zero-temperature case, none
of the derivatives in (\ref{eqn:Projected-0}) is vanishing in
general. The solution to this set of equations is given in
(\ref{eqn:RhoIpm}). Now, by integrating (\ref{eqn:MasterEqB-0})
with the solution (\ref{eqn:RhoIpm}) taken into account and by
going back to the original picture, we obtain the solution to the
master equation (\ref{eqn:Model2-0}):
\begin{align}
\rho(t)
={}&\frac{1}{4}\rho(0)
(1+e^{-\gamma^\beta t}+2e^{-\gamma^\beta t/2}\cos\Omega t)
\nonumber\\
&{}+\frac{1}{4}\sigma_z\rho(0)\sigma_z
(1+e^{-\gamma^\beta t}-2e^{-\gamma^\beta t/2}\cos\Omega t)\nonumber\\
&{}-\frac{1}{4}\rho(0)\sigma_z\left(
\frac{\gamma}{\gamma^\beta}(1-e^{-\gamma^\beta t})
-2ie^{-\gamma^\beta t/2}\sin\Omega t
\right)\nonumber\displaybreak[0]\\
&{}-\frac{1}{4}\sigma_z\rho(0)\left(
\frac{\gamma}{\gamma^\beta}(1-e^{-\gamma^\beta t})
+2ie^{-\gamma^\beta t/2}\sin\Omega t
\right)\nonumber\displaybreak[0]\\
&{}+(1-e^{-\gamma^\beta t})\left(
\frac{\gamma_+}{\gamma^\beta}
\sigma_-\rho(0)\sigma_+
+\frac{\gamma_-}{\gamma^\beta}
\sigma_+\rho(0)\sigma_-
\right).
\label{eqn:SolModel2-0}
\end{align}

At this point, it is worth observing that the above examples
(\ref{eq:ldbdeq3}) and (\ref{eqn:Model2-0}) keep the general
characteristics of the Lindblad master equation, and this
suggests the possibility of applying the procedure to more general cases, which is the subject
of the next section.

\section{Application to More General Master Equations}
\label{sec:general}
So far, we have solved two master equations for finite-level
systems, one at zero and the other at finite temperature. Our approach
essentially consists in writing down a set of equations for
suitable operators through which it is possible to reconstruct the
complete solution of the master equation. In other words, by means
of the resolution of the partial problems given in (\ref{eqn:ElementRhoIEq}) and (\ref{eqn:Projected-0}), it is
possible to give the complete resolution of (\ref{eq:ldbdeq3})
and (\ref{eqn:Model2-0}), respectively. In this section, we are
going to analyze a more general situation and identify a set of
operators which allow us to reproduce the basic steps of the approach
previously exploited. To better understand this point, consider
the following master equation
\begin{subequations}
\label{eq:ldbdeqgen}
\begin{align}
\dot{\rho}(t)
&=-i[H,\rho(t)]\nonumber\\
&{}-\sum_{m,n\in\text{I}}\gamma_{mn}\left(
\frac{1}{2}\{X_{mn}^\dag X_{mn},\rho(t)\}
-X_{mn}\rho(t)X_{mn}^\dag
\right),
\end{align}
where
\begin{equation}
H=\sum_nE_n|n\rangle\langle n|=\sum_nE_nX_{nn},\quad
X_{mn}=|m\rangle\langle n|.
\end{equation}
\end{subequations}
The summations over $m$ and $n$ in the dissipator are taken only over the relevant eigenstates to the interaction with the reservoir, the set of which is symbolically denoted by $\text{I}$\@.
This master equation is not the most general master equation of the Lindblad form, but is the most general one derived from the microscopic Hamiltonian (\ref{eqn:GeneralModel}) under the Born--Markov approximation in the weak-coupling regime \cite{Weiss,ref:MilburnTextbook,ref:KuboTextbook,ref:ScullyTextbook,ref:CarmichaelTextbook,Alicki,Petruccione,Gardiner} or in the van Hove limit \cite{VanHove,VanHove2,ref:QZEdeco}, as far as neither the energy $H^{(0)}$ nor the energy differences
$\omega_{mn}$ in $\text{I}$ is degenerated (see
Appendix~\ref{app:DerivationMasterEq}). These limitations are just
for simplicity, and the procedure works for even more general
cases (for finite-level systems), at least in principle.

Following the procedure presented in the previous section, rearrange the Liouvillian just in the same form as in (\ref{eq:L1}),
\begin{equation}
\dot{\rho}(t)
=A\rho(t)+\rho(t)A^\dag
+\sum_{m,n\in\text{I}}\gamma_{mn}X_{mn}\rho(t)X_{mn}^\dag,
\end{equation}
where
\begin{align}
A&=-iH-\frac{1}{2}\sum_{m,n\in\text{I}}\gamma_{mn}X_{mn}^\dag X_{mn}\nonumber\\
&=-i\sum_n\left(
E_n-\frac{i}{2}\sum_{m\in\text{I}}\gamma_{mn}
\right)X_{nn}.
\end{align}
The transformed operators $X_{mn}(t)$ are easily calculated to be
\begin{align}
X_{mn}(t)
&=e^{-At}X_{mn}e^{At}
\nonumber\\
&=\exp\!\left(
i(E_m-E_n)t+\frac{1}{2}\sum_{\ell\in\text{I}}(\gamma_{\ell  m}-\gamma_{\ell n})t
\right)X_{mn}
\end{align}
and the transformed density matrix $\rho_\text{I}(t)$ satisfies
\begin{align}
\dot{\rho}_\text{I}(t)
&=\sum_{m,n\in\text{I}}\gamma_{mn}X_{mn}(t)\rho_\text{I}(t)X_{mn}^\dag(t)\nonumber\\
&=\sum_{m,n\in\text{I}}\gamma_{mn}
e^{\sum_{\ell\in\text{I}}(\gamma_{\ell  m}-\gamma_{\ell n})t}
X_{mn}\rho_\text{I}(t) X_{mn}^\dag.
\label{eqn:RhoIEqGen}
\end{align}
From this equation, we get
\begin{multline}
\frac{d}{dt}\left(
e^{\sum_{\ell\in\text{I}}(\gamma_{\ell  m}-\gamma_{\ell n})t}
X_{mn}\rho_\text{I}(t)X_{mn}^\dag
\right)\\
=\sum_{n'\in\text{I}}(\Gamma_m)_{nn'}
\left(
e^{\sum_{\ell\in\text{I}}(\gamma_{\ell  m}-\gamma_{\ell n'})t}
X_{mn'}\rho_\text{I}(t) X_{mn'}^\dag
\right),
\label{eqn:EqContractionGen}
\end{multline}
where
\begin{equation}
(\Gamma_m)_{nn'}
=\delta_{nn'}
\sum_{\ell\in\text{I}}(\gamma_{\ell  m}-\gamma_{\ell n})
+\gamma_{nn'},
\label{eq:Gammam}
\end{equation}
and by plugging its solution into (\ref{eqn:RhoIEqGen}), we obtain
\begin{subequations}
\label{eq:MastEqSolution}
\begin{multline}
\rho_\text{I}(t)
=\rho(0)\\
+\sum_{m,n\in\text{I}}\gamma_{mn}\sum_{n'}\left(
\int_0^t e^{\Gamma_ms}ds
\right)_{nn'}X_{mn'}\rho(0)X_{mn'}^\dag,
\label{eq:MastEqSolution1}
\end{multline}
which, in the case wherein the inverse $\Gamma_m^{-1}$ exists, reduces to
\begin{multline}
\rho_\text{I}(t)
=\rho(0)\\
+\sum_{m,n,n'\in\text{I}}\gamma_{mn}
[\Gamma_m^{-1}(e^{\Gamma_mt}-1)]_{nn'}X_{mn'}\rho(0)X_{mn'}^\dag.
\label{eq:MastEqSolution2}
\end{multline}
\end{subequations}
Since the operator $A$ is already diagonal and therefore
\begin{subequations}
\begin{gather}
e^{At}=\sum_ne^{-i\tilde{E}_nt}X_{nn},\quad
e^{A^\dag t}=\sum_ne^{i\tilde{E}_n^*t}X_{nn},\\
\tilde{E}_n=E_n-\frac{i}{2}\sum_{m\in\text{I}}\gamma_{mn},
\end{gather}
\end{subequations}
the solution $\rho(t)=e^{At}\rho_\text{I}(t)e^{A^\dag t}$ easily follows.

From the above procedure for finding the solution to the master equation of the Lindblad form, it is clear that the remaining problem is to calculate the integral of the exponential of the matrix $\Gamma_m$ defined in (\ref{eq:Gammam}).
Such an integral always exists for bounded matrices $\Gamma_m$'s.
Then, the range of effectiveness of this procedure would be evident.
We understand that this procedure is quite useful and effective whenever there are only a few channels through which the system interacts with the reservoir, because in such a case the dimension of the matrix $\Gamma_m$, which is nothing but the number of levels coupled with the reservoir, is small enough to be practically treated, irrespectively of the dimension of the system itself, which may be very large.
On the contrary, as the number of levels interacting with the reservoir increases, the dimension of the matrix $\Gamma_m$ increases linearly and the procedure here expounded becomes less effective.

The above form of the solution (\ref{eq:MastEqSolution}) is very expressive since it separates the part of the evolution associated with the ``quantum jumps,'' $X_{mn'}\rho(0)X_{mn'}^\dag$, and the ``no-jump processes,'' corresponding to $e^{At}\rho(0)e^{A^\dag t}$ \cite{QuantumJump,QuantumJump2}.
It clearly corresponds to the quantum-jump formula given in Refs.~\cite{QuantumJump,QuantumJump2}, but it should be noted that the multiple integrals and the infinite sum of them appearing in such a quantum-jump formal formula, which are usually difficult to carry out directly, are already done in the explicit solution expressed by (\ref{eq:MastEqSolution}).
Moreover, it is worth stressing that the jump contribution is evaluated somehow block by block, hence individualizing a sort of invariant blocks in the Liouville space.
In fact, the matrix $e^{\Gamma_mt}$ is responsible for the evolution of a set of the contractions of the density operator, that is $(X_{m0}\rho X_{m0}^\dag,X_{m1}\rho X_{m1}^\dag,\ldots,X_{m(N-1)}\rho X_{m(N-1)}^\dag)$.
This is well visible for instance from (\ref{eqn:EqContractionGen}).

The master equation (\ref{eq:ldbdeqgen}) is derived under the assumption that there is no degeneracy in the energy difference $\omega_{mn}=E_m^{(0)}-E_n^{(0)}$.
Possible degeneracies in the energy difference would result in a slightly different form of master equation.
In such a case, roughly speaking, the operator $X_{mn}$ in (\ref{eq:ldbdeqgen}) has to be replaced with the sum of operators with the same energy difference and therefore the Liouvillian would get additional off-diagonal terms.
The example (\ref{eqn:Model2}) bears such terms (the second term of the generator).
It is worth stressing that the procedure presented here equally applies to this case as demonstrated in Appendix~\ref{app:SolModel2_HN}, and further, it works even for the most general master equation of the Lindblad form in principle (as far as the number of the decay channels is finite).

\subsection{A Three-Level System at the Zero Temperature}
Finally, it is instructive to consider again the two examples
solved in Sec.~\ref{sec:example}, showing how those solutions are
reproduced through the general formula (\ref{eq:MastEqSolution}).
To this end, it is enough to find the relevant $\Gamma_m$ matrices
defined in (\ref{eq:Gammam}). For the first example
(\ref{eq:ldbdeq3}), only two decay constants are nonvanishing,
$\gamma_{01}=\gamma_{12}=\gamma$, and there are only two relevant
matrices, $\Gamma_m=\gamma\hat{\Gamma}_m$ ($m=0,1$):
\begin{equation}
\hat{\Gamma}_1=
\begin{pmatrix}
1 & 1 & 0 \\
0 & 0 & 1 \\
0 & 0 & 0
\end{pmatrix},\qquad
\hat{\Gamma}_0
=\begin{pmatrix}
0 &  1 &  0 \\
0 & -1 &  1 \\
0 &  0 & -1
\end{pmatrix}.
\label{eqn:GammaModel1}
\end{equation}
They are neither invertible nor diagonalizable.
The exponentials of these matrices are calculated to be
\begin{subequations}
\label{eqn:ExponentialGamma1}
\begin{align}
e^{\Gamma_1t}
&=\begin{pmatrix}
e^{\gamma t} & -1+e^{\gamma t} & -1-\gamma t+e^{\gamma t} \\
0 &  1 &    \gamma t \\
0 &  0 &  1
\end{pmatrix},\displaybreak[0]\\
e^{\Gamma_0t}
&=\begin{pmatrix}
1 & 1-e^{-\gamma t} & 1-e^{-\gamma t}-\gamma te^{-\gamma t} \\
0 & e^{-\gamma t} & \gamma te^{-\gamma t} \\
0 & 0 & e^{-\gamma t}
\end{pmatrix},
\end{align}
\end{subequations}
where $(e^{\Gamma_0t})_{11}$, $(e^{\Gamma_0t})_{12}$, and
$(e^{\Gamma_1t})_{22}$ elements contribute to the solution
(\ref{eq:MastEqSolution1}), which reproduces
(\ref{eqn:RhoIModel1}).

\subsection{A Two-Level System at a Finite Temperature}
For the second example (\ref{eqn:Model2-0}), we have
\begin{equation}
\Gamma_0
=\begin{pmatrix}
0       & \gamma_+\\
\gamma_-&-\gamma
\end{pmatrix},\qquad
\Gamma_1
=\begin{pmatrix}
\gamma  &\gamma_+\\
\gamma_-&0
\end{pmatrix},
\label{eqn:GammaModel2}
\end{equation}
which are invertible and
\begin{subequations}
\label{eqn:ExponentialGamma2}
\begin{align}
e^{\Gamma_0t}
&=\frac{1}{\gamma^\beta}
\begin{pmatrix}
\medskip
\gamma_-e^{-\gamma_+t}+\gamma_+e^{\gamma_-t}&
\gamma_+(e^{\gamma_-t}-e^{-\gamma_+t})\\
\gamma_-(e^{\gamma_-t}-e^{-\gamma_+t})&
\gamma_+e^{-\gamma_+t}+\gamma_-e^{\gamma_-t}
\end{pmatrix},\\
e^{\Gamma_1t}
&=\frac{1}{\gamma^\beta}
\begin{pmatrix}
\medskip
\gamma_+e^{\gamma_+t}+\gamma_-e^{-\gamma_-t}&
\gamma_+(e^{\gamma_+t}-e^{-\gamma_-t})\\
\gamma_-(e^{\gamma_+t}-e^{-\gamma_-t})&
\gamma_-e^{\gamma_+t}+\gamma_+e^{-\gamma_-t}
\end{pmatrix}.
\end{align}
\end{subequations}
The elements $(e^{\Gamma_0t})_{10}$, $(e^{\Gamma_0t})_{11}$, $(e^{\Gamma_1t})_{00}$, and $(e^{\Gamma_1t})_{01}$ are relevant in the solution (\ref{eq:MastEqSolution}), and we obtain (\ref{eqn:SolModel2-0}).

\section{Summary}
\label{sec:conclusion}
In this article, we have presented a general approach to the
resolution of the master equations in the Lindblad form. Through
this procedure, it is possible to solve the Markovian master equation in a compact
form without resort to writing it down as a set of equations among
the matrix elements. The solution is directly given in an operator
form, known as the Kraus representation. Furthermore, one needs to
solve less differential equations than for the matrix elements and
does not need to write down a sparse matrix for the coupled
differential equations among the matrix elements. This difference
is remarkable for relatively simple systems like the first
example, for which the very simple algebra of the operators
involved in the master equation allows us to find the solution in a rather simple operator form.
In fact, the minimal set of the relevant
equations is determined by the structure of a part of the
dissipator, the generator of the evolution of the transformed density operator in (\ref{eqn:RhoIEqGen}), which is responsible for the
so-called
``quantum jumps'' \cite{QuantumJump,QuantumJump2}, and it individualizes a sort of invariant blocks in the Liouville
space. The solution (\ref{eq:MastEqSolution}) to the master
equation (\ref{eq:ldbdeqgen}) is interesting since it clearly
separates the ``quantum jumps'' and the ``no-jump processes.'' A
remarkable point is that, although we demonstrated the present
approach with specific examples, one can solve general Markovian
master equations in principle according to the same idea presented here, as far as the number of the decay
channels is finite.

%--------------------  ACKNOWLEDGEMENTS  --------------------
\section*{Acknowledgements}
The authors thank Professor I. Ohba for discussions.
This work is partly supported by the bilateral Italian--Japanese Projects II04C1AF4E on ``Quantum Information, Computation and Communication'' of the Italian Ministry of Instruction, University and Research, and 15C1 on ``Quantum Information and Computation'' of the Italian Ministry for Foreign Affairs, by the Grant for The 21st Century COE Program ``Holistic Research and Education Center for Physics of Self-Organization Systems'' at Waseda University and the Grants-in-Aid for Scientific Research on Priority Areas ``Dynamics of Strings and Fields'' (No.~13135221) and for Young Scientists (B) (No.~18740250), from the Ministry of Education, Culture, Sports, Science and Technology, Japan, and by the Grants-in-Aid for Scientific Research (C) (Nos.~14540280 and 18540292) from the Japan Society for the Promotion of Science.
Moreover, the authors acknowledge partial support from Universit\`a di Palermo in the context of the bilateral agreement between Universit\`a di Palermo and Waseda University, dated May 10, 2004.

%--------------------  APPENDIXES  ----------------------
\appendix
\section{Derivation of the Lindblad Equations}
\label{app:DerivationMasterEq}
In this appendix, we wish to give possible relationships between the master equations discussed in the text and microscopic Hamiltonians.
Consider a system whose Hamiltonian is diagonalized by a set of nondegenerate eigenstates,
\begin{subequations}
\label{eqn:GeneralModel}
\begin{equation}
H^{(0)}=\sum_nE_n^{(0)}|n\rangle\langle
n|=\sum_nE_n^{(0)}X_{nn},
\end{equation}
and assume that the system interacts with a bosonic reservoir, whose annihilation and creation operators are denoted by $a_{\bm{k}}$ and $a_{\bm{k}}^\dag$, respectively, and
whose Hamiltonian is given by
\begin{equation}
H_\text{R}
=\int d^3\bm{k}\,\omega_{\bm{k}}a_{\bm{k}}^\dag a_{\bm{k}},\qquad
\omega_{\bm{k}}>0.
\label{eqn:BosonicReservoir}
\end{equation}
We consider the general linear interaction between the system and the bosonic reservoir,
\begin{gather}
H_\text{int}=\sum_{m,n\in\text{I}}
(X_{mn}\otimes R^\dag_{mn}+X_{mn}^\dag\otimes R_{mn}),
\displaybreak[0]\\
R_{mn}=\int d^3\bm{k}\,h_{mn}^*(\bm{k})a_{\bm{k}},\quad
X_{mn}=|m\rangle\langle n|,
\end{gather}
\end{subequations}
where the summations over $m$ and $n$ are taken only over the eigenstates actually involved in the interaction with the reservoir. This set is symbolically denoted by $\text{I}$\@.
Note that this interaction Hamiltonian admits counter-rotating terms.
In this appendix, we further assume that there is no degeneracy in the transition frequencies (energy differences)
\begin{equation}
\omega_{mn}=E_m^{(0)}-E_n^{(0)},
\end{equation}
which requires at least $h_{nn}(\bm{k})=0$.

A master equation is then derived from the total Hamiltonian
\begin{equation}
H_\text{tot}=H^{(0)}+H_\text{R}+H_\text{int}
\label{eqn:TotalHamiltonian}
\end{equation}
under the usual Born--Markov approximation in the weak-coupling regime \cite{Weiss,ref:MilburnTextbook,ref:KuboTextbook,ref:ScullyTextbook,ref:CarmichaelTextbook,Alicki,Petruccione,Gardiner} or in the van Hove limit \cite{VanHove,VanHove2,ref:QZEdeco}, and reads (\ref{eq:ldbdeqgen}) \cite{Gardiner,ref:QZEdeco}.
The Hamiltonian $H$ is still diagonalized by the same eigenstates
as $H^{(0)}$ even with the Lamb shifts due to the interaction with the reservoir taken into account.
The decay constants $\gamma_{mn}$ are given by the on-shell form factors with the temperature effect of the reservoir included \cite{ref:QZEdeco},
\begin{equation}
\gamma_{mn}
=[1+N(\omega_{nm})]\Gamma_{mn}(\omega_{nm})
+N(\omega_{mn})\Gamma_{nm}(\omega_{mn}),
\end{equation}
where
\begin{subequations}
\begin{gather}
\Gamma_{mn}(\omega)
=2\pi\int d^3\bm{k}\,|h_{mn}(\bm{k})|^2
\delta(\omega_{\bm{k}}-\omega),\\
N(\omega)=\frac{1}{e^{\beta\omega}-1}.
\end{gather}
\end{subequations}
Note that $\Gamma_{mn}(\omega)=0$ for $\omega<0$ and that the detailed balance condition
\begin{equation}
\gamma_{nm}=e^{-\beta\omega_{mn}}\gamma_{mn}
\end{equation}
holds.

The master equation (\ref{eq:ldbdeqgen}) is not the most general master equation of the Lindblad form.
The characteristic feature of (\ref{eq:ldbdeqgen}) is that the superoperators of the unitary part (the first term of the generator) $\mathcal{H}$ and of the dissipator (the second) $\mathcal{D}$ commute with each other, $[\mathcal{H},\mathcal{D}]=0$.
The other types of master equation are derived in different scaling limits \cite{Gardiner,VanHove2}.

\subsection{Two Qubits Immersed in a Bosonic Reservoir at the Zero Temperature}
\label{app:Model1}
The master equation (\ref{eq:ldbdeq3}) appears in the dynamics of two qubits, say A and B, in interaction with a common bosonic reservoir at temperature $T=0$.
Actually, one can consider the total Hamiltonian (\ref{eqn:TotalHamiltonian}) for the system (two qubits) plus the reservoir, with the Hamiltonians
\begin{subequations}
\begin{align}
H^{(0)}
&=\frac{\Omega}{2}\Sigma_3
+g(\sigma_+^\text{(A)}\sigma_-^\text{(B)}
+\sigma_-^\text{(A)}\sigma_+^\text{(B)}),\displaybreak[0]\\
H_\text{int}
&=\int d^3\bm{k}\,
[h^*(\bm{k})\Sigma_+a_{\bm{k}}
+h(\bm{k})\Sigma_-a^\dag_{\bm{k}}],
\end{align}
\end{subequations}
and (\ref{eqn:BosonicReservoir}), where $\bm{\Sigma}=\bm{\sigma}^\text{(A)}+\bm{\sigma}^\text{(B)}$ is (apart from the normalization) the total spin $\bm{S}$.
$H^{(0)}$ is diagonalized in terms of the total spin $\bm{S}$ ($S=0,1$),
\begin{equation}
H^{(0)}
=\Omega S_3+g[S(S+1)-S_3^2-1]
=\sum_{i=0,1,2,s}E_i^{(0)}|i\rangle\langle i|,
\end{equation}
with
\begin{subequations}
\label{eq:s+t}
\begin{align}
|2\rangle
&=|{\uparrow}\rangle_\text{A}|{\uparrow}\rangle_\text{B},\\
|1\rangle
&=\frac{1}{\sqrt{2}}\,\Bigl(
|{\uparrow}\rangle_\text{A}|{\downarrow}\rangle_\text{B}
+|{\downarrow}\rangle_\text{A}|{\uparrow}\rangle_\text{B}
\Bigr),\\
|0\rangle
&=|{\downarrow}\rangle_\text{A}|{\downarrow}\rangle_\text{B},
\displaybreak[0]\\
|s\rangle
&=\frac{1}{\sqrt{2}}\,\Bigl(
|{\uparrow}\rangle_\text{A}|{\downarrow}\rangle_\text{B}
-|{\downarrow}\rangle_\text{A}|{\uparrow}\rangle_\text{B}
\Bigr),
\end{align}
\end{subequations}
and
\begin{equation}
E_2^{(0)}=\Omega,\quad E_1^{(0)}=g,\quad E_0^{(0)}=-\Omega,\quad E_s^{(0)}=-g.
\end{equation}
The interaction Hamiltonian $H_\text{int}$ is then written as
\begin{equation}
H_\text{int}=\int d{\bm k}\,\sqrt{2}\,h(\bm{k})\,\Bigl(
|0\rangle\langle1|+|1\rangle\langle2|
\Bigr)\,a_{\bm{k}}^\dag+\text{h.c.}
\label{eq:HA+B+bath}
\end{equation}
Notice that the singlet sector is decoupled from the reservoir.
The triplet sector only suffers from dissipation and is described, under the Born--Markov approximation, by the master equation (\ref{eq:ldbdeq3}), provided the decay constants for the two processes, $|2\rangle\to|1\rangle$ and $|1\rangle\to|0\rangle$, given by
\begin{subequations}
\begin{gather}
\gamma_{12}=2\Gamma(\Omega-g),\quad
\gamma_{01}=2\Gamma(\Omega+g),\\
\Gamma(\omega)=2\pi\int d^3\bm{k}\,|h(\bm{k})|^2\delta(\omega_{\bm{k}}-\omega),
\end{gather}
\end{subequations}
are the same, $\gamma_{12}=\gamma_{01}=\gamma$.
It is understood that the energy shifts, say $\delta_i$, due to the interaction have already been included in $E_i$ of (\ref{eqn:SpectrumModel1}).
Actually, we know that
\begin{equation}
E_2=E_2^{(0)}+\delta_2,\
E_1=E_1^{(0)}+\delta_1,\
E_0=E_0^{(0)},\
E_s=E_s^{(0)}.
\end{equation}

\subsection{Spin-Boson Model at a Finite Temperature}
\label{app:Model2}
A master equation of the form
\begin{align}
\dot{\rho}(t)
={}&{-\frac{i}{2}}\Omega[\sigma_z,\rho(t)]
-\gamma_0[\rho(t)-\sigma_z\rho(t)\sigma_z]\nonumber\\
&-\gamma_+\left(
\frac{1}{2}\{\sigma_+\sigma_-,\rho(t)\}
-\sigma_-\rho(t)\sigma_+
\right)\nonumber\\
&-\gamma_-\left(
\frac{1}{2}\{\sigma_-\sigma_+,\rho(t)\}
-\sigma_+\rho(t)\sigma_-
\right),
\label{eqn:Model2}
\end{align}
which reduces to (\ref{eqn:Model2-0}) when $\gamma_0=0$, is derived from the Hamiltonian (\ref{eqn:TotalHamiltonian}) with, for instance \cite{ref:QZEdeco},
\begin{subequations}
\begin{align}
H^{(0)}
={}&\frac{\Omega}{2}\sigma_z,\displaybreak[0]\\
H_\text{int}
={}&\sigma_z\int d^3\bm{k}\,
[h_0^*(\bm{k})a_{\bm{k}}+h_0(\bm{k})a^\dag_{\bm{k}}]\nonumber\\
&{}+\sigma_x\int d^3\bm{k}\,
[h_1^*(\bm{k})a_{\bm{k}}+h_1(\bm{k})a^\dag_{\bm{k}}],
\end{align}
\end{subequations}
and (\ref{eqn:BosonicReservoir}).
Note that, in the above general framework for deriving (\ref{eq:ldbdeqgen}), such terms like the first term of this interaction Hamiltonian $H_\text{int}$, that do not induce transition between the two states of the spin system, are excluded, i.e.~$h_{nn}(\bm{k})=0$.
The second term of the generator of the master equation (\ref{eqn:Model2}) originates from this interaction and is not found in (\ref{eq:ldbdeqgen}).
The decay constants in the generator of (\ref{eqn:Model2}) and (\ref{eqn:Model2-0}) are given by \cite{ref:QZEdeco}
\begin{subequations}
\begin{gather}
\gamma_0=\beta^{-1}\Gamma_0'(0^+),\\
\gamma_+=[1+N(\Omega)]\Gamma_1(\Omega),\quad
\gamma_-=N(\Omega)\Gamma_1(\Omega),
\end{gather}
\end{subequations}
where
\begin{equation}
\Gamma_i(\omega)
=2\pi\int d^3\bm{k}\,|h_i(\bm{k})|^2
\delta(\omega_{\bm{k}}-\omega)\quad(i=0,1).
\end{equation}
Between the two decay constants, the detailed balance condition holds,
\begin{equation}
\gamma_-=e^{-\beta\Omega}\gamma_+.
\label{eqn:DetailedBalance}
\end{equation}

\section{Solving a Master Equation for a Two-Level System}
\label{app:SolModel2_HN}
The master equation (\ref{eqn:Model2}) for a two-level system at a finite temperature is solved as follows.
This is the master equation (\ref{eqn:Model2-0}) when $\gamma_0=0$, and is derived from a spin-boson model: see Appendix~\ref{app:Model2}\@.

First, we arrange the master equation (\ref{eqn:Model2}) into
\begin{align}
\dot{\rho}_\text{I}(t)
={}&\gamma_0\sigma_z\rho_\text{I}(t)\sigma_z\nonumber\\
&{}+\gamma_+\sigma_-\rho_\text{I}(t)\sigma_+e^{-\gamma t}
+\gamma_-\sigma_+\rho_\text{I}(t)\sigma_-e^{\gamma t}\nonumber\\
\equiv{}&(
\gamma_0\mathcal{P}_z+\gamma_+e^{-\gamma t}\mathcal{P}_-+\gamma_-e^{\gamma t}\mathcal{P}_+
)\rho_\text{I}(t),
\label{eqn:M_EqB}
\end{align}
where $\rho_\text{I}(t)$ is defined by (\ref{eqn:IntPic}), i.e.~%
\begin{equation}
\rho(t)=e^{At}\rho_\text{I}(t)e^{A^\dag t}
\equiv e^{\mathcal{A}t}\rho_\text{I}(t)
\end{equation}
with
\begin{subequations}
\begin{gather}
A=-\frac{1}{4}(\gamma^\beta+2\gamma_0)
-\frac{1}{4}(\gamma+2i\Omega)\sigma_z,\\
\gamma^\beta=\gamma_++\gamma_-,\quad
\gamma=\gamma_+-\gamma_-,
\end{gather}
\end{subequations}
and we have introduced superoperators $\mathcal{P}_z$, $\mathcal{P}_\pm$, and $\mathcal{A}$, defined by
\begin{subequations}
\begin{gather}
\mathcal{P}_z\rho=\sigma_z\rho\sigma_z,\quad
\mathcal{P}_-\rho=\sigma_-\rho\sigma_+,\quad
\mathcal{P}_+\rho=\sigma_+\rho\sigma_-,\\
\mathcal{A}\rho=A\rho+\rho A^\dag.
\end{gather}
\end{subequations}
These operators $\mathcal{P}_i$'s satisfy
\begin{subequations}
\label{eq:calP}
\begin{gather}
\mathcal{P}_z^2=1,\quad\mathcal{P}_\pm^2=0,\quad\mathcal{P}_z\mathcal{P}_\pm=\mathcal{P}_\pm=\mathcal{P}_\pm\mathcal{P}_z,\\
\mathcal{P}_-\mathcal{P}_+\mathcal{P}_-=\mathcal{P}_-,\quad\mathcal{P}_+\mathcal{P}_-\mathcal{P}_+=\mathcal{P}_+.
\end{gather}
\end{subequations}
Now define $\tilde\rho_\text{I}(t)=e^{-\gamma_0\mathcal{P}_zt}\rho_\text{I}(t)$, then it satisfies
\begin{equation}
\frac{d}{dt}\tilde{\rho}_\text{I}(t)
=(\gamma_+e^{-\gamma t}\mathcal{P}_-+\gamma_-e^{\gamma t}\mathcal{P}_+)\tilde\rho_\text{I}(t).
\label{eq:rhotilde}
\end{equation}
This equation yields the following equations [notice the relations (\ref{eq:calP})]:
\begin{subequations}
\label{eq:CalPs}
\begin{align}
\frac{d}{dt}
\begin{pmatrix}
\medskip
e^{-\gamma t}\mathcal{P}_-\tilde\rho_\text{I}(t)\\
\mathcal{P}_-\mathcal{P}_+\tilde\rho_\text{I}(t)
\end{pmatrix}
&=\begin{pmatrix}
\medskip
-\gamma&\gamma_-\\
\gamma_+&0
\end{pmatrix}
\begin{pmatrix}
\medskip
e^{-\gamma t}\mathcal{P}_-\tilde\rho_\text{I}(t)\\
\mathcal{P}_-\mathcal{P}_+\tilde\rho_\text{I}(t)
\end{pmatrix},
\\
\frac{d}{dt}
\begin{pmatrix}
\medskip
e^{\gamma t}\mathcal{P}_+\tilde\rho_\text{I}(t)\\
\mathcal{P}_+\mathcal{P}_-\tilde\rho_\text{I}(t)
\end{pmatrix}
&=\begin{pmatrix}
\medskip
\gamma&\gamma_+\\
\gamma_-&0
\end{pmatrix}
\begin{pmatrix}
\medskip
e^{\gamma t}\mathcal{P}_+\tilde\rho_\text{I}(t)\\
\mathcal{P}_+\mathcal{P}_-\tilde\rho_\text{I}(t)
\end{pmatrix},
\end{align}
\end{subequations}
which are easily solved to give
\begin{subequations}
\label{eqn:RhoIpm}
\begin{align}
e^{-\gamma t}\mathcal{P}_-\tilde\rho_\text{I}(t)
=\frac{1}{\gamma^\beta}
\biggl[&
\frac{d}{dt}(e^{\gamma_-t}-e^{-\gamma_+t})\mathcal{P}_-\nonumber\\
&+\gamma_-(e^{\gamma_-t}-e^{-\gamma_+t})\mathcal{P}_-\mathcal{P}_+
\biggr]\rho(0),\\
e^{\gamma t}\mathcal{P}_+\tilde\rho_\text{I}(t)
=\frac{1}{\gamma^\beta}
\biggl[&
\frac{d}{dt}(e^{\gamma_+t}-e^{-\gamma_-t})\mathcal{P}_+\nonumber\\
&+\gamma_+(e^{\gamma_+t}-e^{-\gamma_-t})\mathcal{P}_+\mathcal{P}_-
\biggr]\rho(0).
\end{align}
\end{subequations}
Inserting these results into (\ref{eq:rhotilde}) and integrating its right-hand side, we obtain
\begin{align}
\tilde\rho_\text{I}(t)
=\rho(0)&+\frac{1}{\gamma^\beta}
[
\gamma_+(e^{\gamma_-t}-e^{-\gamma_+t})\mathcal{P}_-\nonumber\\
&\qquad
+(\gamma_+e^{\gamma_-t}+\gamma_-e^{-\gamma_+t}-\gamma^\beta)\mathcal{P}_-\mathcal{P}_+
]\rho(0)\nonumber\\
&+\frac{1}{\gamma^\beta}
[
\gamma_-(e^{\gamma_+t}-e^{-\gamma_-t})\mathcal{P}_+\nonumber\\
&\qquad+(\gamma_-e^{\gamma_+t}+\gamma_+e^{-\gamma_-t}-\gamma^\beta)\mathcal{P}_+\mathcal{P}_-
]\rho(0),
\end{align}
which, together with the relations (\ref{eq:calP}) and $e^{\gamma_0\mathcal{P}_zt}=\cosh\gamma_0t+\mathcal{P}_z\sinh\gamma_0t$, yields
\begin{align}
\rho_\text{I}(t)
={}&(\cosh\gamma_0t+\mathcal{P}_z\sinh\gamma_0t)\rho(0)\nonumber\\
&+\frac{e^{\gamma_0t}}{\gamma^\beta}
[
\gamma_+(e^{\gamma_-t}-e^{-\gamma_+t})\mathcal{P}_-\nonumber\\
&\qquad\qquad
+(\gamma_+e^{\gamma_-t}+\gamma_-e^{-\gamma_+t}-\gamma^\beta)\mathcal{P}_-\mathcal{P}_+
]\rho(0)\nonumber\\
&+\frac{e^{\gamma_0t}}{\gamma^\beta}
[\gamma_-(e^{\gamma_+t}-e^{-\gamma_-t})\mathcal{P}_+\nonumber\\
&\qquad\qquad
+(\gamma_-e^{\gamma_+t}+\gamma_+e^{-\gamma_-t}-\gamma^\beta)\mathcal{P}_+\mathcal{P}_-
]\rho(0).
\end{align}
In order to reach the solution $\rho(t)$, the following relations suffice
\begin{widetext}
\begin{equation}
e^{\mathcal{A}t}\mathcal{P}_-
=e^{-(\gamma_-+\gamma_0)t}\mathcal{P}_-,\qquad
e^{\mathcal{A}t}\mathcal{P}_+
=e^{-(\gamma_++\gamma_0)t}\mathcal{P}_+
\label{eqn:Model2CompBack}
\end{equation}
to give
\begin{align}
\rho(t)
=(\cosh\gamma_0t+\mathcal{P}_z\sinh\gamma_0t)e^{\mathcal{A}t}\rho(0)
&+\frac{1}{\gamma^\beta}
[\gamma_+(1-e^{-\gamma^\beta t})\mathcal{P}_-
+(\gamma_++\gamma_-e^{-\gamma^\beta t}-\gamma^\beta e^{-\gamma_-t})\mathcal{P}_-\mathcal{P}_+
]\rho(0)\nonumber\\
&+\frac{1}{\gamma^\beta}
[\gamma_-(1-e^{-\gamma^\beta t})\mathcal{P}_+
+(\gamma_-+\gamma_+e^{-\gamma^\beta t}-\gamma^\beta e^{-\gamma_+t})\mathcal{P}_+\mathcal{P}_-
]\rho(0).
\end{align}
This reduces to (\ref{eqn:SolModel2-0}) when $\gamma_0=0$.
\end{widetext}

%--------------------  BIBLIOGRAPHY  --------------------

\end{document}